\documentclass[graybox]{svmult}

\usepackage{mathptmx}       % selects Times Roman as basic font
\usepackage{helvet}         % selects Helvetica as sans-serif font
\usepackage{courier}        % selects Courier as typewriter font
\usepackage{type1cm}        % activate if the above 3 fonts are
\usepackage{makeidx}         % allows index generation
\usepackage{graphicx}        % standard LaTeX graphics tool
\usepackage{multicol}        % used for the two-column index
\usepackage[bottom]{footmisc}% places footnotes at page bottom

\makeindex             % used for the subject index
\begin{document}

\title*{Physical properties of 6dF dwarf galaxies}
% Use \titlerunning{Short Title} for an abbreviated version of
% your contribution title if the original one is too long
\author{Jean Michel Gomes and Polychronis Papaderos}
% Use \authorrunning{Short Title} for an abbreviated version of
% your contribution title if the original one is too long
\institute{Jean Michel Gomes and Polychronis Papaderos \at CAUP - Centro de
  Astrof\'isica da Universidade do Porto \newline Rua das Estrelas, 4150-762
  Porto, Portugal \newline \email{jean@astro.up.pt and papaderos@astro.up.pt}}
%\and Polychronis Papaderos \at CAUP - Centro de Astrof\'isica da Universidade do Porto \newline Rua da%s Estrelas, 4150-762 Porto, Portugal \newline \email{papaderos@astro.up.pt}}
%
% Use the package "url.sty" to avoid
% problems with special characters
% used in your e-mail or web address
%
\maketitle

\vspace*{-20mm} \abstract{Spectral synthesis is basically the
  decomposition of an observed spectrum in terms of the superposition
  of a base of simple stellar populations of various ages and
  metallicities, producing as output the star formation and chemical
  histories of a galaxy, its extinction and velocity dispersion.
 \newline\indent The {\sc starlight} code provides one of the most
 powerful spectral synthesis tools presently available. We have
 {applied} this code to the entire Six-Degree-Field Survey (6dF)
 sample of nearby star-forming galaxies, selecting dwarf galaxy
 candidates with the goal of
\begin{itemize}
\item 
  deriving the age and metallicity of their stellar populations
\item and creating a database with the physical properties of our sample galaxies
  together with the FITS files of pure emission line spectra (i.e. the
  observed spectra after subtraction of the best-fitting synthetic
  stellar spectrum).
\end{itemize}
Our results yield a good qualitative and quantitative agreement with
previous studies based on the Sloan Digital Sky Survey (SDSS).
However, an advantage of 6dF spectra is that they are taken within a
twice as large fiber aperture, much reducing aperture effects in
studies of nearby dwarf galaxies.}

\section{Introduction}
\label{sec:1}
We have entered in a new era with the availability of high-quality
spectroscopic data bases for large galaxy samples, such as the Sloan
Digital Sky Survey (SDSS) and Six-Degree Field (6dF) Surveys. The
combination of a wide set of synthetic and observed stellar libraries
with meanwhile much refined population synthesis codes permits us to
significantly improve our understanding of the formation and evolution
of galaxies.

One such publicly available code for the derivation of physical
properties of galaxies is {\sc starlight} \cite{REF4}.  The main
output from the model, $M(\lambda)$, is a linear combination of
$N_\star$ Simple Stellar Populations (SSPs) of different age and
metallicity. This study uses Bruzual \& Charlot SSP models
\cite{REF1}.  The basic equation is:

\begin{equation}
\frac{M(\lambda)}{M(\lambda_0)} = \sum_{j=1}^{N_\star} x_{j,\lambda_0}
b_{j,\lambda_0}(\lambda) r(\lambda) \otimes G(v_\star,\sigma_\star)
\end{equation}

\noindent where, $M(\lambda_0)$ is the flux of the best-fitting model
at the normalization wavelength $\lambda_0$, $x_{j,\lambda_0}$ is the
j$^{\textrm{th}}$ SSP flux contribution at $\lambda_0$ to the modeled
spectrum, $b_{j,\lambda_0}(\lambda)$ is the j$^{\textrm{th}}$ SSP
spectrum normalized at $\lambda_0$, $r(\lambda)$ is the extinction law
(Cardelli, Clayton, \& Mathis ~\cite{REF3}). $G(v_\star,\sigma_\star)$
is a Gaussian folding function that is used to take into account the
stars' velocity dispersion $\sigma_\star$ and the systemic velocity
$v_\star$.  

The principle by-product of this equation is the Star Formation
History (SFH) encoded in the population vector expressed in terms of
$x_{j,\lambda_0}$. Therefrom we derive the luminosity-weighted mean
stellar age ($\langle \log t_\star \rangle_L = \sum_{j=1}^{N_\star}
x_{j,\lambda_0} \log t_j$) and metallicity ($\langle Z_\star \rangle_L =
\sum_{j=1}^{N_\star} x_{j,\lambda_0} Z_j$), where $t_j$ and $Z_j$ are
the age and metallicity of the j$^{\textrm{th}}$ SSP and secondary
quantities, such as the Mass-to-Light (M2L) ratio in the $K$, $H$ and
$J$ filters, which together with photometry, allows us to estimate the
total stellar mass ($M_\star$) of galaxies.

\section{The 6dF Sample}
\label{sec:2}
We have studied the entire final data release of the 6dF Survey
\cite{REF5}, which comprises spectra and $K$, $H$ and $J$ 2MASS
photometry for 136 304 galaxies with a mean redshift of $0.053$, much
lower than that of SDSS. The spectra were obtained in two observations
using separate V and R gratings, that together yield a resolution of
$\sim 1000$ over the spectral range $4000$--$7500$ \AA \textrm{ } and
a signal-to-noise ratio $\sim 10$ per pixel.  After a series of tests,
we decided to model the V part of the spectrum only due to its better
calibration, which is vital for a reliable derivation of the SFH with
{\sc starlight}. However, emission lines from both parts of the
spectrum have been used to distinguish between star-forming galaxies
and AGNs, based on the diagnostic emission-line ratios: [OIII]$\lambda
5007 / $H$\beta$ and [NII]$\lambda 6583 / $H$\alpha$ in BPT diagrams
\cite{REF0}.

\subsection{Dwarf Galaxy Candidates in the 6dF}
\label{subsec:2}
\begin{figure}[t]
%\sidecaption[t]
% Use the relevant command for your figure-insertion program
% to insert the figure file.
% For example, with the option graphics use
\includegraphics[scale=.55,viewport=-20 300 530 540]{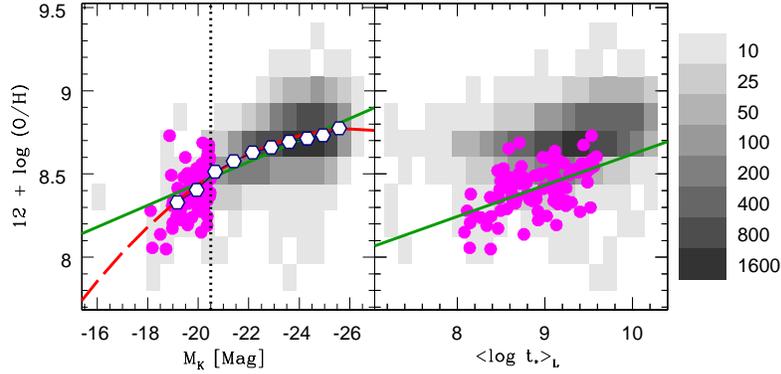}
%
% If no graphics program available, insert a blank space i.e. use
%\picplace{5cm}{2cm} % Give the correct figure height and width in cm
%
%\caption{Please write your figure caption here}
\caption{Comparison of the subsample of 90 star-forming dwarf galaxies
  (filled magenta circles) with the entire High Quality (HQ) sample of
  star-forming galaxies (7938 objects) in the 6dF. The grey-scale
  distribution in the right bar depicts the number of galaxies in each
  bin.  {\bf Left panel:} Absolute $K$ magnitude vs gas-phase
  metallicity.  The vertical dotted line marks the absolute magnitude
  range of $-20.5 \leq M_K \leq -13.5$ adopted for the extraction of
  the dwarf galaxy sample. The solid and dashed lines show a
  linear and polynomial fit to the median trend shown with white
  hexagons. {\bf Right panel:} Luminosity-weighted mean stellar age vs
  nebular metallicity for the HQ 6dF sample. A linear fit to the dwarf
  galaxy subsample (solid line) suggest a slight steepening of the
  stellar age vs metallicity relation for low-mass galaxies.}
\label{fig:1}       
\end{figure}

After selecting spectral fits with a percentage deviation from input spectra
of less than 5\%, we extracted a sub-sample of $\sim$ 20 000 galaxies.  
This is the high-quality (HQ) sample, out of which we selected 116 dwarf galaxies
by using a ``soft'' criterion that is based on the absolute $K$
magnitude ($-20.5 \leq M_K \leq -13.5$; Fig. \ref{fig:1}). 
%According to the BPT classification scheme, 90 of these galaxies are
%star-forming. 
90 of these galaxies are star-forming.

\section{Results}
\label{sec:3}

We computed the nebular metallicity using the N2 index (N2 $\equiv$
log [NII]$\lambda 6583 / $H$\alpha$) following the parametrization of
Pettini \& Pagel \cite{REF7}. In Fig. \ref{fig:1} we show the absolute
magnitude $M_K$ vs the nebular metallicity (right panel). The dotted
line marks the absolute magnitude range adopted for the selection of
dwarf galaxies. The luminosity-weighted mean stellar age vs
gas-phase metallicity for the dwarf galaxy candidates is shown on the
right panel. 
We can see that dwarf galaxy candidates span $\sim$ 2 dex with respect to
$\langle \log t_\star \rangle_L$ with systems among them harboring old stellar 
populations being predominantly metal-rich and {\it vice versa}.

\begin{figure}[t]
\sidecaption[t]
% Use the relevant command for your figure-insertion program
% to insert the figure file.
% For example, with the option graphics use
\includegraphics[scale=.55,viewport=-10 21 530 550]{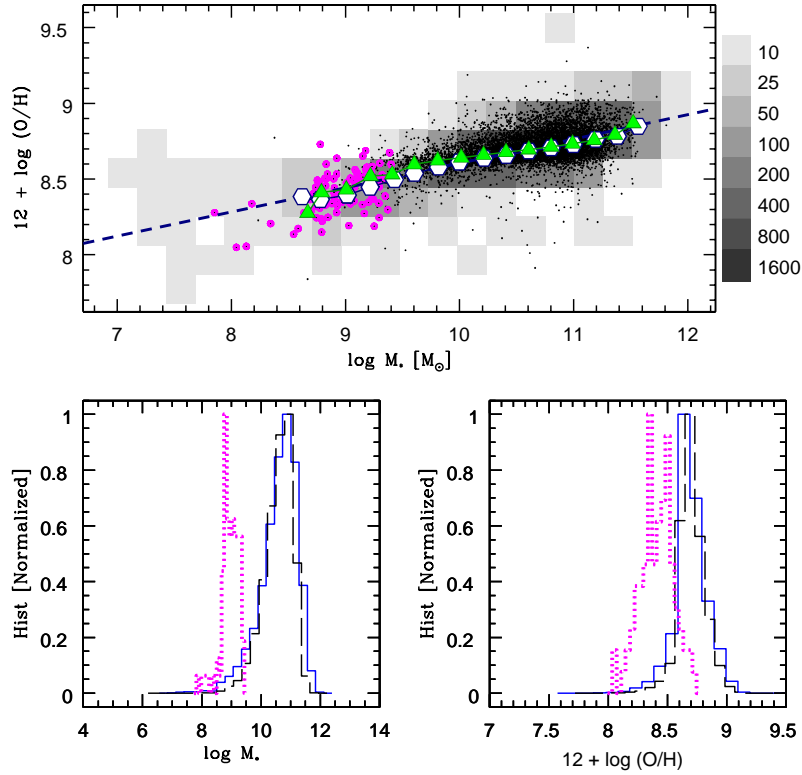}
%
% If no graphics program available, insert a blank space i.e. use
%\picplace{5cm}{2cm} % Give the correct figure height and width in cm
%
%\caption{Please write your figure caption here}
\caption{{\bf Top panel:} Stellar mass vs nebular metallicity (MZ)
  relation for the 6dF HQ sample (black dots) and the dwarf galaxy
  sub-sample (filled circles).  For comparison, we show the SDSS DR7
  sample with a grey-scale distribution, representing the number of
  galaxies in each pixel (vertical bar). The median trend is shown
  with green triangles and white hexagons for the 6dF and SDSS DR7
  sample, respectively. Linear fits to both 6dF and SDSS galaxies
  (dashed line) are practically identical, independently confirming
  previous results (e.g.: Tremonti et al. 2004).  {\bf Bottom Panels:}
  Histograms with the stellar mass (left) and the gas-phase
  metallicity (right) of the HQ and dwarf galaxy sample from the 6dF,
  and from the SDSS (dotted, dashed and solid lines, respectively).  }
\label{fig:2}       
\end{figure}

An important outcome from this study is outlined in Fig. \ref{fig:2} where
we show the stellar mass--nebular metallicity (MZ) relation, and the
stellar mass and gas-phase metallicity distributions for the entire HQ
sample and the sub-sample of the 90 dwarf galaxy candidates. The MZ relation,
as derived from the SDSS DR7 data is included for comparison. 
The dwarf galaxy candidates, whose
median stellar mass and nebular metallicity is by, respectively,
$\sim$ 2 and $\sim$ 0.3 dex lower than the values determined for the
entire HQ data set, follow the overall trend found for star-forming galaxies,
independently corroborating previous studies
(\cite{REF6,REF8,REFa}, among others).
A follow-up analysis and discussion of the MZ relation, based on 6dF data 
is planned in \cite{REF9}.

%\vspace*{-5mm}
\begin{acknowledgement}
J. M. Gomes is supported by a Post-Doctoral grant, funded by FCT/MCTES
(Portugal) and POPH/FSE (EC) and P. Papaderos is supported by Ciencia
2008 Contract, funded by FCT/MCTES (Portugal) and POPH/FSE (EC).
\end{acknowledgement}

\vspace*{-5mm}
%%%%%%%%%%%%%%%%%%%%%%%% referenc.tex %%%%%%%%%%%%%%%%%%%%%%%%%%%%%%
% sample references
% %
% Use this file as a template for your own input.
%
%%%%%%%%%%%%%%%%%%%%%%%% Springer-Verlag %%%%%%%%%%%%%%%%%%%%%%%%%%
%
% BibTeX users please use
% \bibliographystyle{}
% \bibliography{}

\begin{thebibliography}{99.}%

% Journal article
%\bibitem{sj1} Hamburger, C.: Quasimonotonicity, regularity and duality for nonlinear systems of partial differential equations. Ann. Mat. Pura. Appl. \textbf{169}, 321--354 (1995)

\bibitem{REFa} Asari et al. 2009, MNRAS, 396L, 71
\bibitem{REF0} Baldwin J. A., Phillips M. M., Terlevich R., PASP, 93, 5
\bibitem{REF1} Bruzual G., Charlot S. 2003, MNRAS, 344, 1000
\bibitem{REF3} Cardelli et al. 1989, ApJ, 345, 245
\bibitem{REF4} Cid Fernandes et al. 2005, MNRAS, 358, 363
%\bibitem{REF5} Finlator \& Dav\'e 2008
\bibitem{REF5} Jones et al. 2009, MNRAS, 399, 683
\bibitem{REF6} Lequeux J. et al. 1979, A\&A, 80, 155
\bibitem{REF7} Pettini M., Pagel B. 2004, MNRAS, 348L, 59
\bibitem{REF8} Tremonti C. A. et al. 2004, ApJ, 613, 898
\bibitem{REF9} Gomes, J.-M. \& Papaderos, P. 2011, in prep.

\end{thebibliography}
%
%\biblstarthook{}

\end{document}